\documentclass[reprint,
superscriptaddress,
amsmath,amssymb,
aps,
pra,
floatfix,
]{revtex4-1}
\usepackage{graphicx}
\usepackage{float}
\usepackage{hyperref}
\hypersetup{colorlinks=true,linkcolor=blue,citecolor=blue,urlcolor=blue}

\usepackage{siunitx}
\usepackage{apjfonts}

\newcommand{\LINE}[1]{{#1}}
\newcommand{\ION}[2]{{{#1}~\textsc{#2}}}
\newcommand{\PETRAIII}{{PETRA~\textsc{III}}}

\newcommand{\ambit}{\texttt{AMBiT}}

\begin{document}

	
	\title{{High Resolution Photoexcitation Measurements Exacerbate\\ the Long-standing {Fe}~{\normalfont\textsc{xvii}} Oscillator Strength Problem}}

	
	\author{Steffen K\"uhn}\thanks{These two authors have contributed equally.\\ \href{mailto:steffen.kuehn@mpi-hd.mpg.de}{steffen.kuehn@mpi-hd.mpg.de};  \href{mailto:chintan@mpi-hd.mpg.de}{chintan@mpi-hd.mpg.de}}
	\affiliation{Max-Planck-Institut f\"ur Kernphysik, Saupfercheckweg 1, 69117 Heidelberg, Germany}%
	\affiliation{Heidelberg Graduate School of Fundamental Physics, Ruprecht-Karls-Universit\"at Heidelberg, Im Neuenheimer Feld 226, 69120 Heidelberg, Germany}
	
	\author{Chintan Shah}\thanks{These two authors have contributed equally.\\ \href{mailto:steffen.kuehn@mpi-hd.mpg.de}{steffen.kuehn@mpi-hd.mpg.de};  \href{mailto:chintan@mpi-hd.mpg.de}{chintan@mpi-hd.mpg.de}}
	\affiliation{NASA/Goddard Space Flight Center, 8800 Greenbelt Rd, Greenbelt, Maryland 20771, USA}%
	\affiliation{Max-Planck-Institut f\"ur Kernphysik, Saupfercheckweg 1, 69117 Heidelberg, Germany}%
	
	\author{Jos\'e R. Crespo L\'opez-Urrutia}%
	\affiliation{Max-Planck-Institut f\"ur Kernphysik, Saupfercheckweg 1, 69117 Heidelberg, Germany}%
	
	\author{Keisuke Fujii}%
	\affiliation{Department of Mechanical Engineering and Science, Graduate School of Engineering, Kyoto University, Kyoto 615-8540, Japan}%
	
	\author{Ren\'e Steinbr\"ugge}
	\affiliation{Deutsches Elektronen-Sychrotron DESY, Notkestra{\ss}e 85, 22607 Hamburg, Germany}%
	
	\author{Jakob Stierhof}
	\affiliation{Dr. Karl Remeis-Sternwarte, Sternwartstra{\ss}e 7, 96049 Bamberg, Germany}%
	
	\author{Moto Togawa}%
	\affiliation{Max-Planck-Institut f\"ur Kernphysik, Saupfercheckweg 1, 69117 Heidelberg, Germany}%
	
	\author{Zolt\'an Harman}
	\author{Natalia S. Oreshkina}
	\affiliation{Max-Planck-Institut f\"ur Kernphysik, Saupfercheckweg 1, 69117 Heidelberg, Germany}%
	
	\author{Charles Cheung}
	\affiliation{Department of Physics and Astronomy, University of Delaware, Newark, Delaware 19716, USA}
	
	\author{Mikhail G. Kozlov}
	\affiliation{Petersburg Nuclear Physics Institute of NRC "Kurchatov Institute", Gatchina 188300, Russia}
	\affiliation{St. Petersburg Electrotechnical University "LETI", Prof. Popov Str. 5, St. Petersburg, 197376, Russia}
	
	\author{Sergey G. Porsev}
	\affiliation{Petersburg Nuclear Physics Institute of NRC "Kurchatov Institute", Gatchina 188300, Russia}
	\affiliation{Department of Physics and Astronomy, University of Delaware, Newark, Delaware 19716, USA}
	
	\author{Marianna S. Safronova}
	\affiliation{Department of Physics and Astronomy, University of Delaware, Newark, Delaware 19716, USA}
	\affiliation{Joint Quantum Institute, National Institute of Standards and Technology and the University of Maryland, Gaithersburg, Maryland 20742, USA}
	
	\author{Julian C. Berengut}
	\affiliation{School of Physics, University of New South Wales, Sydney, New South Wales 2052, Australia}%
	\affiliation{Max-Planck-Institut f\"ur Kernphysik, Saupfercheckweg 1, 69117 Heidelberg, Germany}%
	
	\author{Michael Rosner}%
	\affiliation{Max-Planck-Institut f\"ur Kernphysik, Saupfercheckweg 1, 69117 Heidelberg, Germany}%
	
	\author{Matthias Bissinger}
	\affiliation{Erlangen Centre for Astroparticle Physics (ECAP), Erwin-Rommel-Stra{\ss}e 1, 91058 Erlangen, Germany}
	\affiliation{Dr. Karl Remeis-Sternwarte, Sternwartstra{\ss}e 7, 96049 Bamberg, Germany}%
	
	\author{Ralf Ballhausen}
	\affiliation{Dr. Karl Remeis-Sternwarte, Sternwartstra{\ss}e 7, 96049 Bamberg, Germany}%
	
	\author{Natalie Hell}
	\affiliation{Lawrence Livermore National Laboratory, 7000 East Avenue, Livermore, California 94550, USA}%
	
	\author{SungNam Park}
	\author{Moses Chung}
	\affiliation{Ulsan National Institute of Science and Technology, 50 UNIST-gil, 44919 Ulsan, South Korea}
	
	\author{Moritz Hoesch}
	\author{J\"orn Seltmann}
	\affiliation{Deutsches Elektronen-Sychrotron DESY, Notkestra{\ss}e 85, 22607 Hamburg, Germany}%
	
	\author{Andrey S. Surzhykov}
	\affiliation{Physikalisch-Technische Bundesanstalt, Bundesalle 100, 38116 Braunschweig, Germany}
	\affiliation{Braunschweig University of Technology, Universit{\"a}tsplatz 2, 38106 Braunschweig, Germany }%
	
	\author{Vladimir A. Yerokhin}
	\affiliation{Peter the Great St.Petersburg Polytechnic University, 195251 St.~Petersburg, Russia}%
	
	\author{J\"orn Wilms}
	\affiliation{Dr. Karl Remeis-Sternwarte, Sternwartstra{\ss}e 7, 96049 Bamberg, Germany}%
	
	\author{F. Scott Porter}
	\affiliation{NASA/Goddard Space Flight Center, 8800 Greenbelt Rd, Greenbelt, MD 20771, USA}%
	
	\author{Thomas St\"ohlker}
	\affiliation{Institut f\"ur Optik und Quantenelektronik, Friedrich-Schiller-Universit\"at Jena,  Max-Wien-Platz 1, 07743 Jena, Germany}%
	\affiliation{Helmholtz-Institut Jena, Fr\"obelstieg 3, 07743 Jena, Germany}%
	\affiliation{GSI Helmholtzzentrum f\"ur Schwerionenforschung, Planckstra{\ss}e 1, 64291 Darmstadt, Germany}%
	
	\author{Christoph H. Keitel}
	\author{Thomas Pfeifer}
	\affiliation{Max-Planck-Institut f\"ur Kernphysik, Saupfercheckweg 1, 69117 Heidelberg, Germany}%
	
	\author{Gregory V. Brown}
	\affiliation{Lawrence Livermore National Laboratory, 7000 East Avenue, Livermore, CA 94550, USA}%
	
	\author{Maurice A. Leutenegger}
	\affiliation{NASA/Goddard Space Flight Center, 8800 Greenbelt Rd, Greenbelt, MD 20771, USA}%
	
	\author{Sven Bernitt}
	\affiliation{Max-Planck-Institut f\"ur Kernphysik, Saupfercheckweg 1, 69117 Heidelberg, Germany}%
	\affiliation{Institut f\"ur Optik und Quantenelektronik, Friedrich-Schiller-Universit\"at Jena,  Max-Wien-Platz 1, 07743 Jena, Germany}%
	\affiliation{Helmholtz-Institut Jena, Fr\"obelstieg 3, 07743 Jena, Germany}%
	\affiliation{GSI Helmholtzzentrum f\"ur Schwerionenforschung, Planckstra{\ss}e 1, 64291 Darmstadt, Germany}%

	
	\date{\today}

	
	\begin{abstract}
			For more than 40 years, most astrophysical observations and laboratory studies of two key soft X-ray diagnostic $2p-3d$ transitions, \LINE{3C} and \LINE{3D}, in {Fe~\scriptsize{XVII}} ions {found} oscillator strength ratios $f(\mathrm{\LINE{3C}})/f(\mathrm{\LINE{3D}})$ disagreeing with theory, but uncertainties had precluded definitive statements on this much studied conundrum. Here, we resonantly excite these lines using synchrotron radiation at PETRA III, and reach, at a millionfold lower photon intensities, a 10 times higher spectral resolution, and 3 times smaller uncertainty than earlier work. Our final result of $f(\mathrm{\LINE{3C}})/f(\mathrm{\LINE{3D}}) = 3.09(8)(6)$ supports many of the earlier clean astrophysical and laboratory observations, while departing by five sigmas from our own newest large-scale \emph{ab initio} calculations, and excluding all proposed explanations, including those invoking nonlinear effects and population transfers.
	\end{abstract}

	\maketitle
	

	\begin{figure*}
		\includegraphics[width=\textwidth]{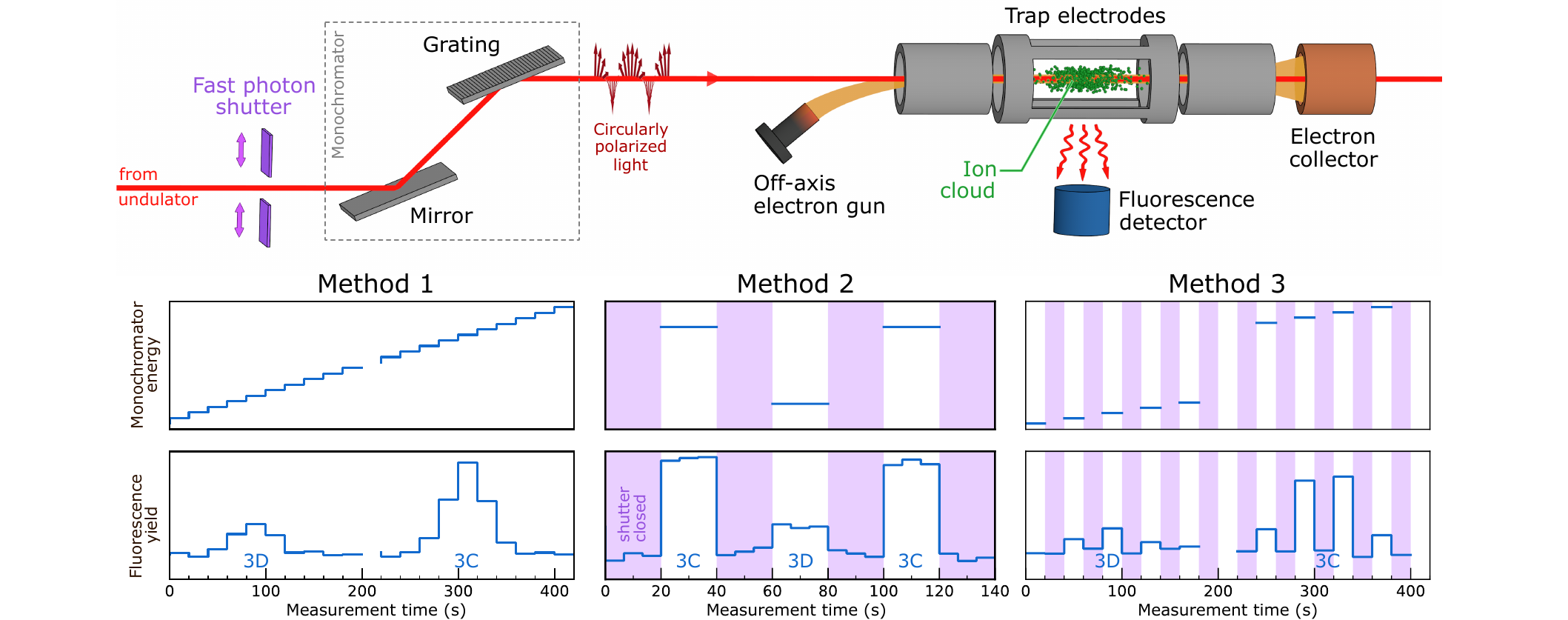}
		\caption{%
			{(Top) Experimental setup: an electron beam (orange) aimed at the trap center produces \ION{Fe}{xvii} ions, which are then resonantly excited by a monochromatic photon beam (red). Subsequent X-ray fluorescence is registered by a silicon drift detector.
			(Bottom) Fluorescence yield and photon energy vs. time with three different methods: (1) line scans with open photon shutter; (2,3) by closing it at each step (purple areas), we subtract the electron-beam-induced background at (2) each line center and (3) scanning over a one-full-width-at-half-maximum range.}
			}
		\label{fig:setup}
	\end{figure*}

	
	Space X-ray observatories, such as \emph{Chandra} and \emph{XMM-Newton}, resolve $L$-shell transitions of iron dominating the spectra of many hot astrophysical objects~\cite{pfk2003,chd2000,behar2001,mrd2001}. Some of the brightest lines arise from \ION{Fe}{xvii}~(Ne-like iron) around 15~\AA: the resonance line \LINE{3C}~($[(2p^5)_{1/2}\,3d_{3/2}]_{J=1}\rightarrow[2p^6]_{J=0}$) and the intercombination line \LINE{3D}~($[(2p^5)_{3/2}\,3d_{5/2}]_{J=1}\rightarrow[2p^6]_{J=0}$).
	{Appearing over a broad range of plasma temperatures and densities, they are crucial for diagnostics of electron temperatures, elemental abundances, ionization conditions, velocity turbulences, and opacities~\cite{par1973,doron2002,xpb2002,brickhouse2005,saf2011,kem2014,pzw2012,bnl2014,beiersdorfer2018,nagayama2019,gu2019}.}
	However, for the past four decades, their observed intensity ratios persistently disagree with advanced plasma models, diminishing the utility of high resolution X-ray observations.
	{Several experiments} using electron beam ion trap (EBIT) and tokamak devices {have} scrutinized plausible astrophysical and plasma physics explanations as well as the underlying atomic theory~\cite{bbl1998,beiersdorfer2001,bbc2001,bbb2002,bbg2004,bbc2006,brown2012,shah2019}, {but also revealed} clear departures from predictions {while broadly agreeing with} astrophysical observations~\cite{bbb2002,bbg2004,gu2009}.
	This has fueled a long-lasting controversy on the cause being a lack of understanding of astrophysical plasmas, or inaccurate atomic data.

	A direct probe of these lines using an EBIT at the Linac Coherent Light Source (LCLS) X-ray free-electron laser (XFEL) found again their oscillator strength ratio $f(\mathrm{\LINE{3C}})/f(\mathrm{\LINE{3D}})$ to be lower than predicted, but close to astrophysical observations~\cite{sbr2012}.
	{This highlighted difficulties with oscillator strengths calculations in many-electron systems~\cite{safronova2001,cpr2002,lpb2006,che2007,chen2011,gu2009,sbr2012,slt2015}.}
	{Nonetheless, at} the high peak brilliance of the LCLS XFEL, nonlinear excitation dynamics~\cite{ock2014,oreshkina2016x} or nonequilibrium time evolution~\cite{lbl2015} might have affected {the result of \citet{sbr2012}}.
	An effect of resonance-induced population transfer between \ION{Fe}{xvi} and \ION{Fe}{xvii} ions was also postulated~\cite{wu2019} since the \ION{Fe}{xvi} line \LINE{C}~($ \left[ (2p^5)_{1/2} 3s3d_{3/2} \right ]_{J=1/2} \to \left[ 2p^6 3s \right]_{J=1/2} $) appeared blended with the \ION{Fe}{xvii} line \LINE{3D}.
	A recent semiempirical calculation~\cite{mba2017} reproduces the LCLS results~\cite{sbr2012} by fine-tuning relativistic couplings and orbital relaxation effects, but its validity has been disproved~\cite{wpe2017}.
	
	{
	In this Letter, we report on new measurements of resonantly excited \ION{Fe}{xvi} and \ION{Fe}{xvii} with a synchrotron source {at} tenfold improved spectral resolution and  millionfold lower peak photon flux than in~\cite{sbr2012}, suppressing nonlinear dynamical effects~\cite{ock2014,lbl2015,li2017} and undesired ion population transfers~\cite{wu2019}.
	We also carry out improved large-scale calculations using three different {advanced} approaches~\cite{kozlov2015,Fischer2019,kahl2019ambit}, all showing a five-sigma departure from our experimental results.
	}	
	
	
	\begin{figure*}
		\includegraphics[width=0.9\textwidth]{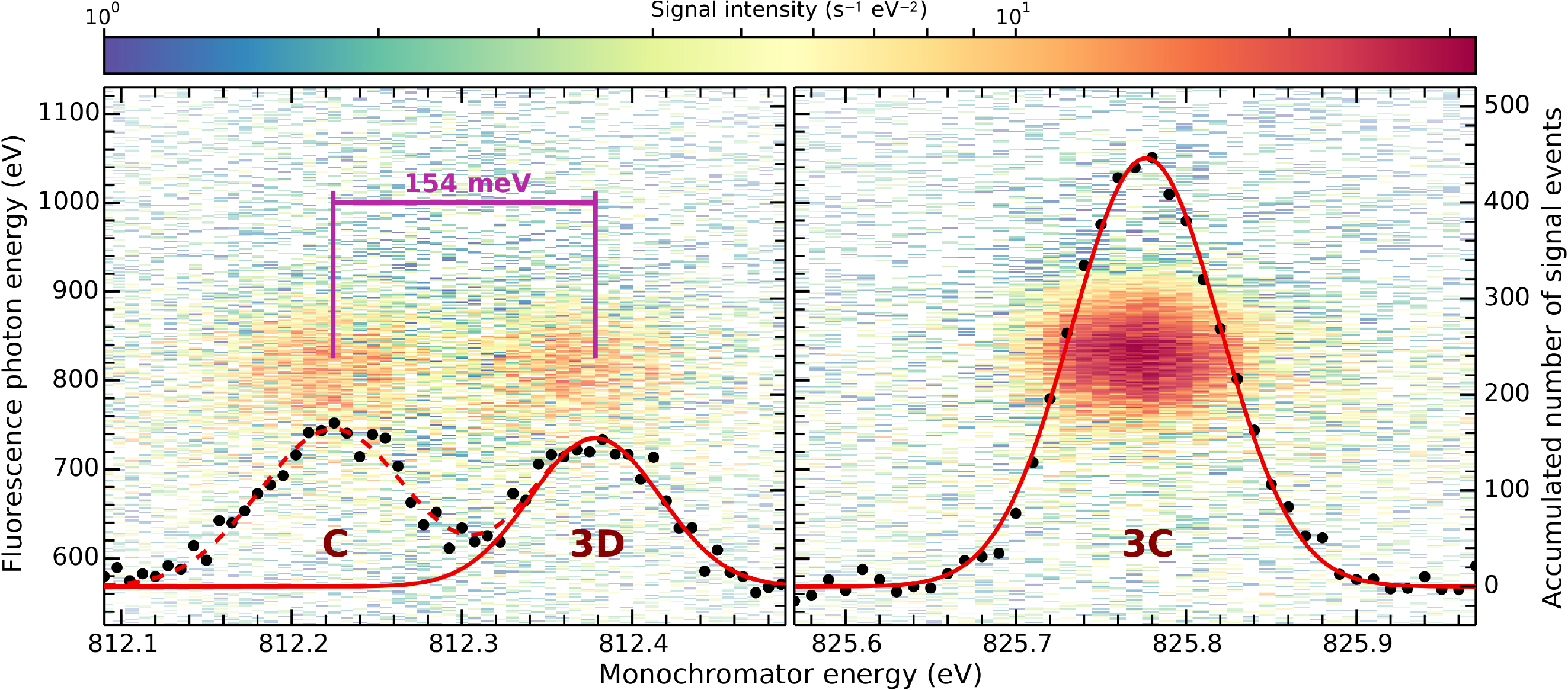}
		\caption{
			Fluorescence photon yield and energy vs~excitation-photon energy for the \ION{Fe}{xvi} \LINE{C}, \ION{Fe}{xvii} \LINE{3C}, and \LINE{3D} transitions recorded by a silicon-drift detector. Black dots: total fluorescence within a 50-eV region of interest. Red solid lines: Fits to \LINE{3C} and \LINE{3D}. Red dashed line: fit to \LINE{C}.
		}
		\label{fig:2D}
	\end{figure*}

	
	We used the compact PolarX-EBIT~\cite{micke2018}, in which a monoenergetic electron beam emitted by an off-axis cathode (see Fig.~\ref{fig:setup}) is compressed by a magnetic field. 
	{At the trap center, it collides with a beam of iron-pentacarbonyl molecules, dissociating them, and producing highly charged Fe ions with a relative abundance of {Fe~\scriptsize{XVI}} to {Fe~\scriptsize{XVII}} close to unity. }
	These ions stay radially confined by the negative space charge of the $\sim${2}-{mA}, {1610}-{eV} ($\approx3$ times the \ION{Fe}{xvi} ionization potential) electron beam, and axially by potentials applied to surrounding electrodes. 

	Monochromatic, circularly polarized photons from the P04 beamline~\cite{viefhaus2013} at the \PETRAIII{} synchrotron photon source enter through the electron gun, irradiate the trapped ions, and exit through the collector aperture. They can resonantly excite X-ray transitions on top of the strong electron-induced background due to ionization, recombination, and excitation processes. 
	A side-on-mounted {energy-resolving windowless silicon drift photon detector (SDD) equipped with a 500-nm thin aluminum filter} registers these emissions.
	%

	By scanning the P04 monochromator between 810 and 830~eV, we excite the \ION{Fe}{xvii} lines \LINE{3C} and \LINE{3D}, as well as the \ION{Fe}{xvi} lines \LINE{B} ($ \left[ (2p^5)_{1/2} (3s3d_{3/2})_{1} \right ]_{J=1/2} \to \left[ 2p^6 3s \right]_{J=1/2} $) and \LINE{C}. They are also nonresonantly excited by electron-impact as the electron beam energy is well above threshold~\cite{bbc2006,shah2019}.
	This leads to a {strong, nearly constant} X-ray background at the same energies as the photoexcited transitions {but independent of the exciting photon-beam energy}.
	{In our earlier work ~\cite{sbr2012}, we rejected this background by detecting the fluorescence in time coincidence with the sub-picosecond-long LCLS pulses of $\approx10^{11}$ photons each at 120/s.
	Due to much longer and weaker 50-ps-long pulses ($\approx10^{3}$ photons, $6\times10^{6}$/s repetition rate) at P04 and the limited time resolution of SDD, we could not use the coincidence method, and reached a signal-to-background ratio of only $\sim$5\%.
	To improve this, we used a shutter to cyclically turn on and off the P04 photon beam.
	}
	
	{    
	Using a 50-$\mu$m slit width, we reached a resolving power of ${E}/{\Delta E}\approx 10000$, ten to fifteen times higher than that of~\emph{Chandra} and~\emph{XMM-Newton} grating spectrometers~\cite{den2001reflection,canizares2005chandra}, and {tenfold} that of our previous experiment~\cite{sbr2012} (See Supplemental Material (SM)).
	We find a separation of \LINE{3C} from \LINE{3D} of $\Delta{E}_{3C-3D}={13.398(1)}~{\mathrm{eV}}$, and resolve for the first time the \ION{Fe}{xvii} \LINE{3D} line from the \ION{Fe}{xvi} \LINE{C} one at $\Delta{E}_{3D-C}={154.3(1.3)}~{\mathrm{meV}}$.
	This gives us the \LINE{3C}/\LINE{3D} intensity ratio without having to infer a contribution of \ION{Fe}{xvi} line \LINE{C} (in \cite{sbr2012} still unresolved) from the intensity of the well-resolved \ION{Fe}{xvi}  \LINE{A} line.
	Thereby, we largely reduce systematic uncertainties and exclude the resonance-induced population transfer mechanism~\cite{wu2019} that may have affected the LCLS result~\cite{sbr2012}.
	}

	{We apply three different methods (Fig.~\ref{fig:setup}) to systematically measure the \LINE{3C}/\LINE{3D} oscillator strength ratio.}
	In {method~1}, we did not {operate} the photon shutter; instead, we repeatedly scanned the lines \LINE{C} and \LINE{3D} (812.0 -- \SI{812.5}{\eV}), as well as \LINE{3C} (825.5 -- \SI{826.0}{\eV}), in both cases using scans of 100 steps with 20-s exposure each (see Fig.~\ref{fig:2D}).
	{The SDD fluorescence signal was integrated} over a 50-eV wide photon-energy region of interest (ROI) comprising \LINE{3C}, \LINE{3D}, and \LINE{C}, and recorded {while scanning} the incident photon energy.
	By fitting Gaussians {to the scan result}, we obtain line positions, widths, and yields, modeling the electron-impact background as a smooth linear function~\cite{shah2019}.
	The ratio of \LINE{3C} and \LINE{3D} areas is then proportional to the oscillator strength ratio~\cite{oreshkina2016x}.
	However, {given the low 5\% signal-to-background ratio and long measurement times, changes in the background cause systematic uncertainties}.
	In {method~2}, we fixed the monochromator energy to the respective centroids of \LINE{C}, \LINE{3D}, and \LINE{3C} found with {method 1}, and cyclically opened and closed the shutter for equal periods of \SI{20}{\second} to determine the background.
	The background-corrected fluorescence yields at the line peaks were multiplied with the respective linewidths from {method~1}, to obtain the \LINE{3C}/\LINE{3D} ratio.
	Still, slow monochromator shifts from the selected positions could affect the results.
	{To address this, in {method~3}, we scanned across the FWHM of \LINE{C}, \LINE{3D}, and \LINE{3C} in 33 steps with on-off exposures of \SI{20}{\second}, which reduced the effect of possible monochromator shifts.}
	After background subtraction, we fit Gaussians to the lines of interest fixing their widths to values from {method~1}.

\begin{figure}
		\includegraphics[width=\columnwidth]{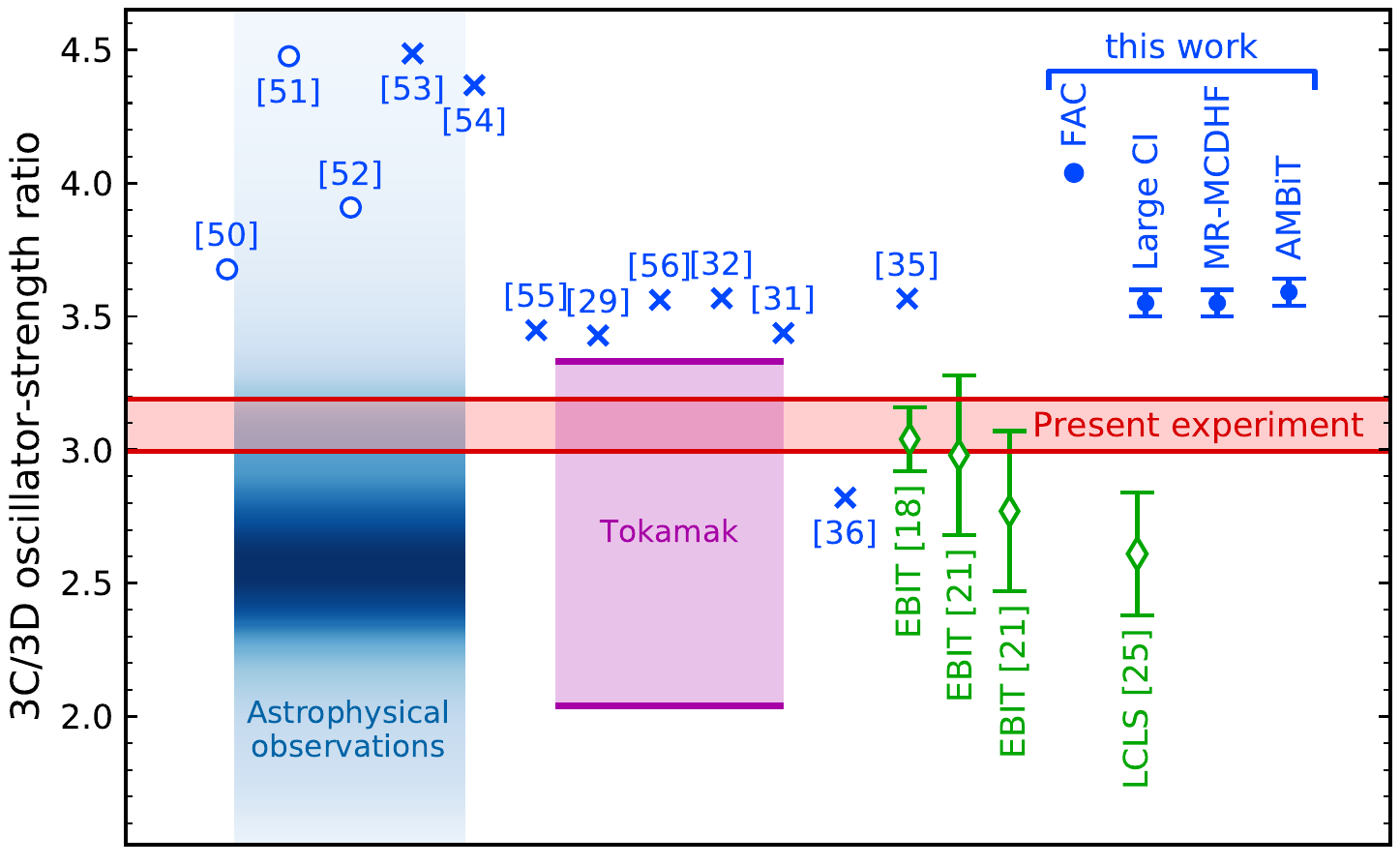}
		\caption{
			Present experimental \LINE{3C}/\LINE{3D} ratios compared with previous predictions and experiments. 
			Red band: combined results of the three different methods. 		%
			Blue open circles: values from databases~\cite{NIST_ASD,kaastra1996,fsb2012}. 
			Blue crosses: predictions~\cite{bhatia1992,chen2003,dong2003,che2007,ock2014,jonsson2014,slt2015,mba2017,wu2019}. Note that the validity of theory~\cite{mba2017} has been disputed~\cite{wpe2017}.
			Blue solid circles: present FAC~\cite{gu2011fac}, large-scale CI~\cite{kozlov2015}, MR-MCDHF~\cite{Fischer2019}, and AMBiT~\cite{kahl2019ambit} calculations. 
			Blue band: observed line ratios in astrophysical sources~\cite{blake1965,mckenzie1980,mrd2001,behar2001,xpb2002,ness2003}, {with color shades} coding the distribution of values weighted by their reported accuracies. 
			Purple band: spread of tokamak results~\cite{bbg2004}. 		%
			Open green diamonds: previous EBIT results~\cite{bbc2001,bbc2006,sbr2012}. 
			{Note that the spread seen in various astrophysical sources and in tokamak {in part arises} from insufficient removal of {Fe~\scriptsize{XVI}}~\LINE{C} line contamination of {Fe~\scriptsize{XVII}}~\LINE{3D} line, at varying {Fe~\scriptsize{XVI}}/{Fe~\scriptsize{XVII}} abundance ratios~\cite{doron2002,bbg2004}; nonlinear dynamical effects~\cite{ock2014,lbl2015,li2017} (see main text) may have reduced the LCLS ratio~\cite{sbr2012} .}
		}
		\label{fig:ratio}
	\end{figure}

	All three methods share systematic uncertainties caused by energy-dependent filter transmission and detector efficiency~($\sim$1\%) and by the incident photon beam flux variation~($\sim$2\%).
	Additionally, for {method~1}, we estimate systematic uncertainties from background~($\sim$1.2\%) and ROI selection~($\sim$2.7\%).
	In {method~2}, possible monochromator shifts from (set) line centroids and widths taken from {method~1} cause a systematic uncertainty of $\sim$3.5\%.
	Analogously, for {method~3}, we estimate a $\sim$3\% uncertainty due to the use of linewidth constraints from {method~1}.
	{The weighted average of all three methods is $f(\mathrm{3C})/f(\mathrm{3D}) = 3.09(8)_{\mathrm{sys}}(6)_{\mathrm{stat}}$, see Fig.~\ref{fig:ratio} (see SM for individual ratio and uncertainties).}
	Note that the circular polarization of the photon beam does not affect these results, since \LINE{3C} and \LINE{3D} (both ${\Delta}{J}=1$) share the same angular emission characteristics~\cite{balashovbook,rudolph2013x,shah2015,sas2018}.

	Calculations using a density-matrix approach by~\citet{oreshkina2016x,ock2014}, and the time-dependent collisional-radiative model of~\citet{lbl2015}, pointed to a possible nonlinear response of the excited upper state populations in~\cite{sbr2012}, reducing the observed oscillator strength ratio~\cite{li2017}, {which would depend} on photon pulse parameters, like intensity, duration, and spectral distribution. 
	It has been estimated that peak intensities above 10$^{12}$ W/cm$^2$ would give rise to nonlinear effects~\cite{ock2014,li2017}. 
	Fluctuations of the self-amplified spontaneous emission process at LCLS can conceivably generate some pulses above that threshold. 
	At P04, we estimate a peak intensity of $\approx$10$^{5}$ W/cm$^2$, more than 6 orders of magnitude below that threshold (see SM).
	This could explain why the \LINE{3C}/\LINE{3D} ratio in~\citet{sbr2012} is in slight disagreement with the present result. 
	Nonetheless, our experiment validates the main conclusion of that work with reduced uncertainty.
	Our work implies that future experiments at ultra-brilliant light sources should take possible nonlinear effects into account.

	In the present work, we also carried out relativistic calculations using a very-large-scale configuration interaction (CI) method, correlating all ten electrons, including Breit and quantum electrodynamical~\cite{tupitsyn2016quantum} corrections.
	We implemented a message passing interface (MPI) version of the CI code from ~\cite{kozlov2015} to increase the number of configurations to over {230,000}, saturating the computation in all possible numerical parameters. 
	Basis sets of increasing size are used to check for convergence, with all orbitals up to $12sp17dfg$ included in the largest version (the contributions of $n>12\,\,sp$ orbitals are negligible).
	We start with all possible single and double excitations from the $2s^2 2p^6$, $2s^2 2p^5 3p$  even and
	$2s^2 2p^5 3s$, $2s^2 2p^5 3d$, $2s 2p^6 3p$, $2s^2 2p^5 4d$, $2s^2 2p^5 5d$ odd configurations, correlating eight electrons.
	We separately calculate triple excitations and fully correlate the $1s^2$ shell, and also included dominant quadruple ones, finding them negligible.
	The line strengths $S$ and \LINE{3C}/\LINE{3D} oscillator strength ratio after several computation stages are summarized in the SM to illustrate the small effect of all corrections.
	Theoretical uncertainties are estimated based on the variance of results from the smallest to largest runs, size of the various effects, and small variances in the basis set construction.
	We verified that the energies of all 18 states considered, counted from the ground state, agree with the National Institute for Standards and Technology~\cite{NIST_ASD} database well within the experimental uncertainty of 0.05\%.
	The theoretical 3C-3D energy difference of 13.44~eV is in agreement with the experiment to 0.3\%.

	{We also carried out entirely independent large-scale calculations using the multireference multiconfiguration Dirac-Hartree-Fock (MR-MCDHF) approach~\cite{Fischer2019} with up to 1.2 million configurations.}
	First, the $2s^2 2p^5 3s$, $2s^2 2p^5 3d$ and $2s 2p^6 3p$ $J=1$ levels were used as reference states to generate the list of configuration state functions with single and double exchanges from all occupied orbitals up to $12spdfghi$.
	Virtual orbitals were added in a layer-by-layer manner.
	Subsequently, the role of triple excitations was studied by the CI method.
	In a second step, the multireference list was extended to include all $J=1$ odd parity states, generated from the Ne-like ground state by single and double electron exchanges.
	Monitoring the convergence of the results for the addition of layers of virtual orbits, we arrive at an oscillator strength ratio of 3.55(5), and to a 3C-3D energy splitting of 13.44(5)~eV.
	Another full-scale CI calculation with more than a million configurations was carried out in the particle-hole formalism using AMBiT~\cite{kahl2019ambit}, agreeing well with the other theoretical results.
	Full details of all calculations can be found in SM.
	We emphasize that there are no other known quantum mechanical effects or numerical uncertainties to consider within the CI and MCDHF approaches.
	With modern computational facilities and MPI codes, we have shown that all other contributions are negligible at the level of the quoted theoretical uncertainties.
	The significant improvements in experimental and theoretical precision reported here have only further deepened this long-standing problem.
	{This work on the possibly so far most intensively studied many-electron ion in experiment and theory, finally demonstrates convergence of the dedicated atomic calculations on all possible parameters, excluding an incomplete inclusion of the correlation effects as potential explanation of this puzzle.}

	{
	Our result is the presently most accurate on the \LINE{3C}/\LINE{3D} oscillator strength ratio.
	Its excellent resolution {suggests promising direct determinations of the natural linewidth. 
	They depend on the Einstein $A$ coefficients, hence, on the oscillator strengths~\cite{Weisskopf1997}}. 
	Thus, future accurate measurements of {individual} natural linewidths of \LINE{3C} and \LINE{3D} not only would test theory more stringently than their oscillator strength ratio does, but also deliver accurate oscillator strengths.
	}

	{
	Moreover, \LINE{3C} and \LINE{3D} with their, among many transitions, strong absorption and emission rates can also dominate the Planck and Rosseland mean opacity of hot plasmas~\cite{rogers1994,seaton1994,beiersdorfer1996observation}. 
	Therefore, an accurate determination of their oscillator strengths may help elucidating the iron opacity issue~\cite{bailey2007,bnl2014,nagayama2019}, if, e.~g., Rosseland mean opacity models~\cite{fontes2015,pain2017} were found to use predicted oscillator strengths also in departure from experiments.
	Our result exposes in simplest dipole-allowed transitions of {Fe~\scriptsize{XVII}} a far greater issue, namely, the persistent problems in the best approximations in use, and calls for renewed efforts in further developing the theory of many-electron systems.
	}
	
	
	%
	Shortcomings of low-precision atomic theory for $L$-shell ions {had already emerged} in the analysis of high resolution \emph{Chandra} and \emph{XMM-Newton} data~\cite{bbb2002,bbg2004,netzer2004,gu2009,beiersdorfer2018}.
	Similar inconsistencies were recently found in the high resolution $K$-shell X-ray spectra of the Perseus cluster recorded with the~\textit{Hitomi}~microcalorimeter~\cite{hitomi2016,hitomi2018}.
	{Moreover, recent opacity measurements~\cite{bnl2014,nagayama2019} {have} highlighted serious inconsistencies {in} the opacity models used to describe the interiors of stars, which have to rely on calculated oscillator strengths}.

	{All this shows} that the actual accuracy and reliability of the opacity and turbulence velocity diagnostics are still uncertain, and with them, the modeling of hot astrophysical and high-energy density plasmas.
	The upcoming X-ray observatory missions~\textit{XRISM}~\cite{xrism2018} and ~\textit{Athena}~\cite{barret2016} will require improved and quantitatively validated modeling tools {for maximizing their scientific harvest}.
	{Thus, benchmarking atomic theory in the laboratory is vital}.
	As for the long-standing \ION{Fe}{xvii} oscillator strength problem, our results may be {immediately used to semiempirically correct spectral models of astrophysical observations}.

	
	
	\begin{acknowledgments}
		Financial support was provided by the Max-Planck-Gesellschaft (MPG) and Bun\-des\-mi\-ni\-ste\-ri\-um f{\"u}r Bildung und Forschung (BMBF) through Project No. 05K13SJ2.
		Work by C.S. was supported by the Deutsche Forschungsgemeinschaft (DFG) Project No. 266229290 and by an appointment to the NASA Postdoctoral Program at the NASA Goddard Space Flight Center, administered by Universities Space Research Association under contract with NASA.
		Work by LLNL was performed under the auspices of the U.S. Department of Energy under Contract No. DE-AC52-07NA27344 and supported by NASA grants.
		M.A.L. and F.S.P. acknowledge support from NASA's Astrophysics Program.
		The work of M.G.K. and S.G.P. was supported by the Russian Science Foundation under Grant No. 19-12-00157.
		The theoretical research was supported in part through the use of Information Technologies resources at the University of Delaware, specifically the high-performance Caviness computing cluster.
		The work of C.C. and M.S.S. was supported by U.S. NSF Grant No. PHY-1620687.
		Work by UNIST was supported by the National Research Foundation of Korea (No. NRF-2016R1A5A1013277).
		J.C.B. acknowledges support from the Alexander von Humboldt Foundation.
		{We acknowledge DESY (Hamburg, Germany), a member of the Helmholtz Association HGF, for the provision of experimental facilities. Parts of this research were carried out at PETRA III.}
	\end{acknowledgments}

	
%


\widetext
\clearpage


	\begin{center}
		\textbf{\large High resolution Photoexcitation Measurements Exacerbate\\ the Long-standing {Fe}~{\normalfont\textsc{xvii}} Oscillator Strength Problem: Supplemental Material}
	\end{center}
	
	\setcounter{equation}{0}
	\setcounter{figure}{0}
	\setcounter{table}{0}
	\setcounter{page}{1}
	\makeatletter
	\renewcommand{\theequation}{S\arabic{equation}}
	\renewcommand{\thefigure}{S\arabic{figure}}
	\renewcommand{\bibnumfmt}[1]{[S#1]}
	\renewcommand{\citenumfont}[1]{S#1}
	\renewcommand{\thetable}{S\arabic{table}}
	
	\section{Experiment and Data Analysis}

\begin{table}[ht]
	\centering
	\caption{Comparison between experimental values and theoretical predictions of the $3C/3D$ oscillator strength ratio and relative line energy positions achieved within this work.}
	
	\begin{tabular}{lcccc}
		\hline\hline
		\multicolumn{1}{l}{} & \multicolumn{1}{c}{Experiment} & \multicolumn{1}{c}{CI}       & \multicolumn{1}{c}{MCDHF}    & \multicolumn{1}{c}{\ambit}   \\
		\hline
		3C/3D oscillator strength ratio &$3.09(8)_{\mathrm{sys}}(6)_{\mathrm{stat}}$& 3.55(5)& 3.55(5)  & 3.59(5)   \\
		Energy 3C (eV)                    &            & 825.67   & 825.88(5)& 825.923 \\
		Energy 3D (eV)                     &            & 812.22   & 812.44(5)& 812.397 \\
		$\Delta$Energy 3C-3D (eV)                        & 13.398(1)  & 13.44(5) & 13.44(5) & 13.526  \\
		$\Delta$Energy 3D-C (eV)                         & 0.1543(13) &          &          & \\
		Natural linewidth 3C (meV)    &  & 14.74(3)  & 14.75(3) & 14.90\\
		Natural linewidth 3D (meV)     &  & 4.02(5)  & 4.01(6) & 4.04\\
		
		\hline\hline      
	\end{tabular}
\end{table}

\subsection{Individual methods data and their uncertainities}

\begin{figure}[!ht]
	\includegraphics[width=0.6\textheight]{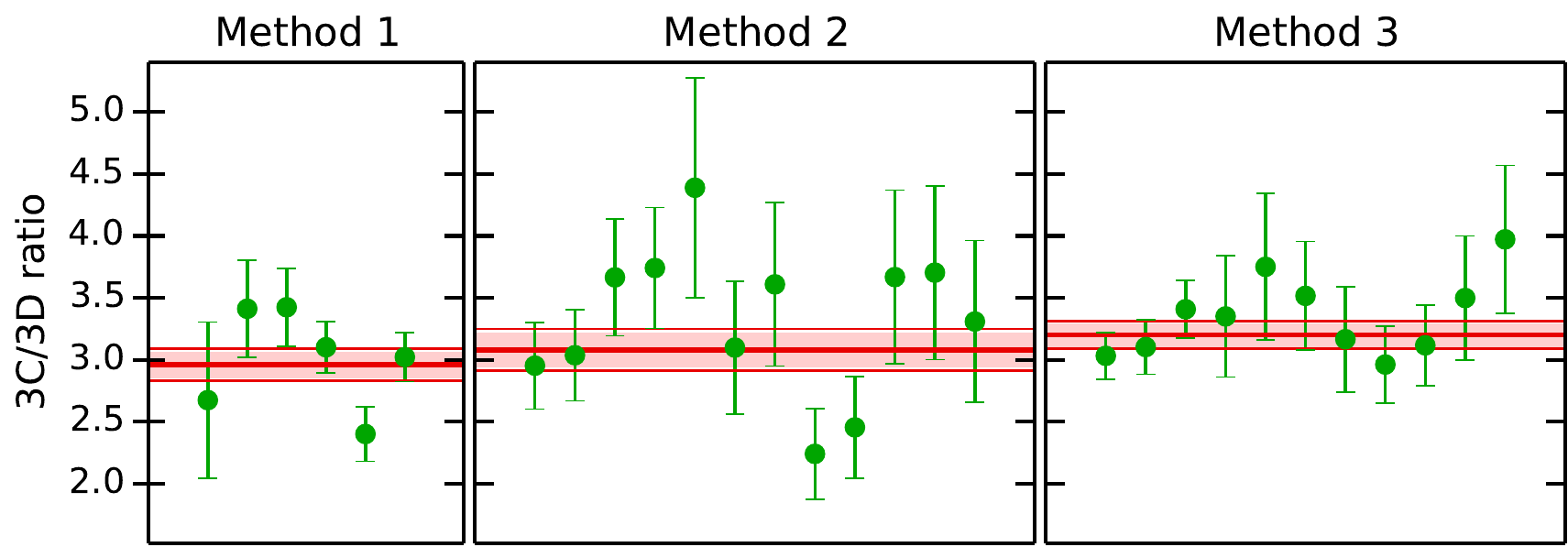}
	\caption{Experimental \LINE{3C}/\LINE{3D} ratio for the three different methods. The individual measurements are shown as solid green circles. The weighted mean experimental values and associated 1-$\sigma$ statistical and systematic uncertainties are indicated as red band.}
	\label{fig:methods}
\end{figure}

\begin{table}[htbp]
	\centering
	\caption{$3C/3D$ oscillator strength ratios obtained from three diffferent measurement methods and their statistical and systematic uncertainties.}
	\begin{tabular}{lccc}
		\hline\hline
		& Method 1 & Method 2 & Method 3 \\
		\hline
		$3C/3D$ oscillator strength ratio & 2.960 & 3.080 & 3.210 \\
		\hline\hline
		\multicolumn{4}{c}{Uncertainty Budget} \\
		\hline
		Statistical & 0.106 & 0.140 & 0.095 \\
		\hline
		Systematics due to: &       &       &  \\
		(1) ROI width selection on 2D histogram (Fig.~2 of the main paper) & 0.030 &       &  \\
		(2) ROI centroid selection on 2D histogram & 0.044 &       &  \\
		(3) Filter transmission and efficiency of the detector & 0.030 & 0.031 & 0.032 \\
		(4) Time-dependent background variation due to the electron-impact excitation & 0.036 &       &  \\
		(5) Monochromator shifts in the (set) energy position &       & 0.092 &  \\
		(6) Linewidth constraints in Gaussian fits &       &       & 0.048 \\
		\hline
		Total systematic uncertainty & 0.071 & 0.097 & 0.058 \\
		\hline\hline
		Total (statistical + systematic) uncertainties & 0.127 & 0.170 & 0.111 \\
		\hline\hline
		Common systematics for all three methods: &       &       &  \\
		Flux variation of the incident photon beam at P04/PETRA III & \multicolumn{3}{c}{0.0618} \\
		\hline
		&       &       &  \\
		\textbf{Final $3C/3D$ oscillator strength ratio} & \multicolumn{3}{c}{\textbf{3.09 $\pm$ 0.08$_\mathrm{stat.}$ $\pm$ 0.06$_\mathrm{sys.}$}} \\
		&       &       &  \\
		\hline\hline
	\end{tabular}%
	\label{tab:systematics}%
\end{table}%

\newpage

\subsection{Resolving  {Fe~\scriptsize{XVII}} 3D and {Fe~\scriptsize{XVI}} C lines}

\begin{figure}[!ht]
	\includegraphics[width=0.6\textheight]{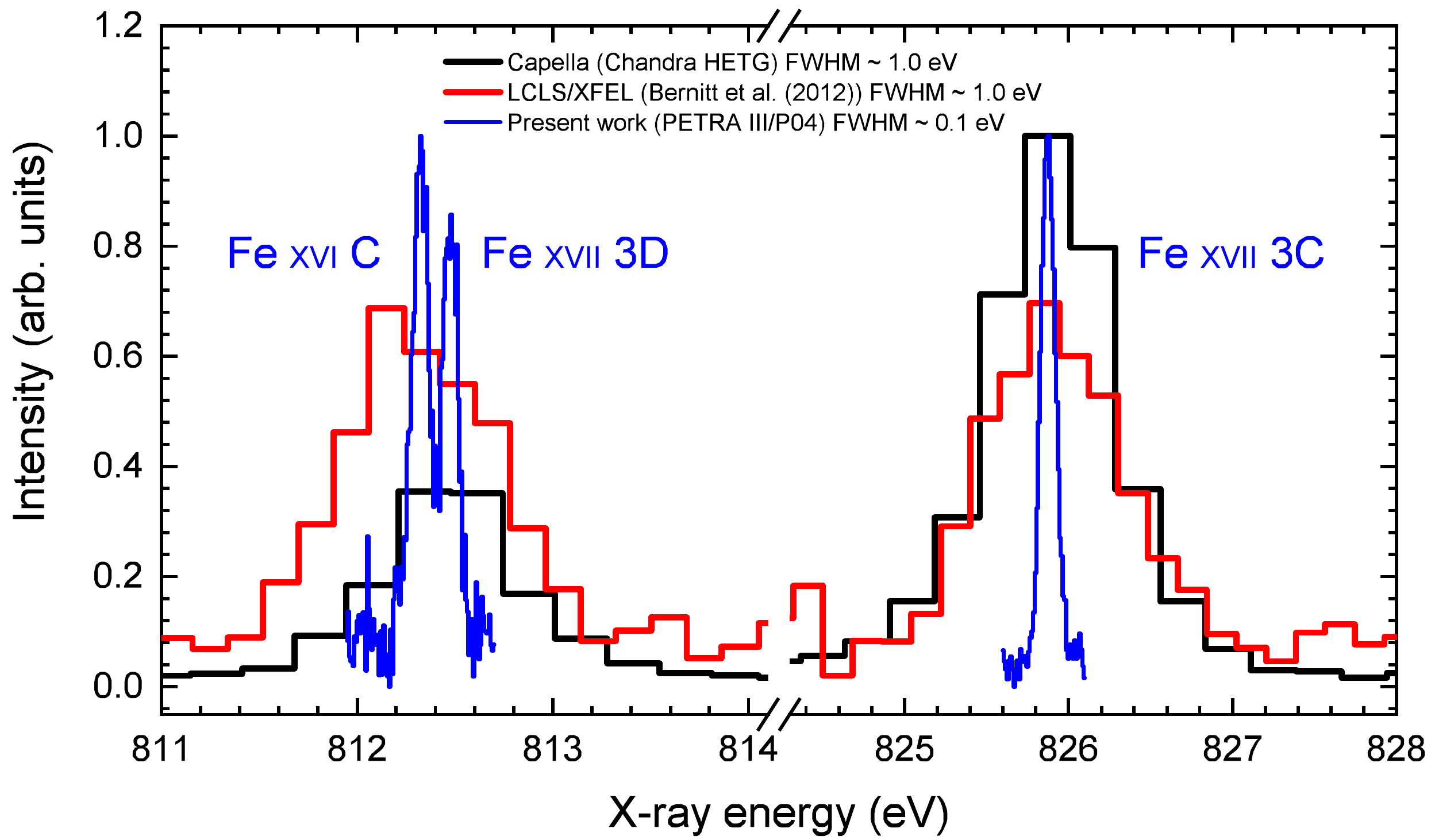}
	\caption{Spectral resolution comparison between the present work at \PETRAIII/P04, our previous work at LCLS/XFEL~\cite{S_sbr2012}, and the high resolution~\textit{Chandra} High Energy Transmission Grating (HETG)~\cite{S_chd2000} spectrum of Capella [ObsId: 1103]. In this work, for the first time, we have resolved previously-blended \ION{Fe}{xvi} \LINE{C} from \ION{Fe}{xvii} \LINE{3D} line, which are 154.3(1.3) meV apart from each other. This enables us to obtain the line intensity ratio of \ION{Fe}{xvii} \LINE{3C} and \LINE{3D} without having to subtract the contribution of the \ION{Fe}{xvi} \LINE{C} line, in contrast to all other previous works. Moreover, this has largely reduced the systematic uncertainties and eliminated the need for taking resonance-induced population transfer~\cite{S_wu2019} into account, which may have affected the accuracy of our LCLS work~\cite{S_sbr2012}.}
	\label{fig:resolution}
\end{figure}

\subsection{Comparison between experimental data, theoretical predictions and astrophysical observations}
\begin{figure}[!ht]
	\includegraphics[width=0.8\textwidth]{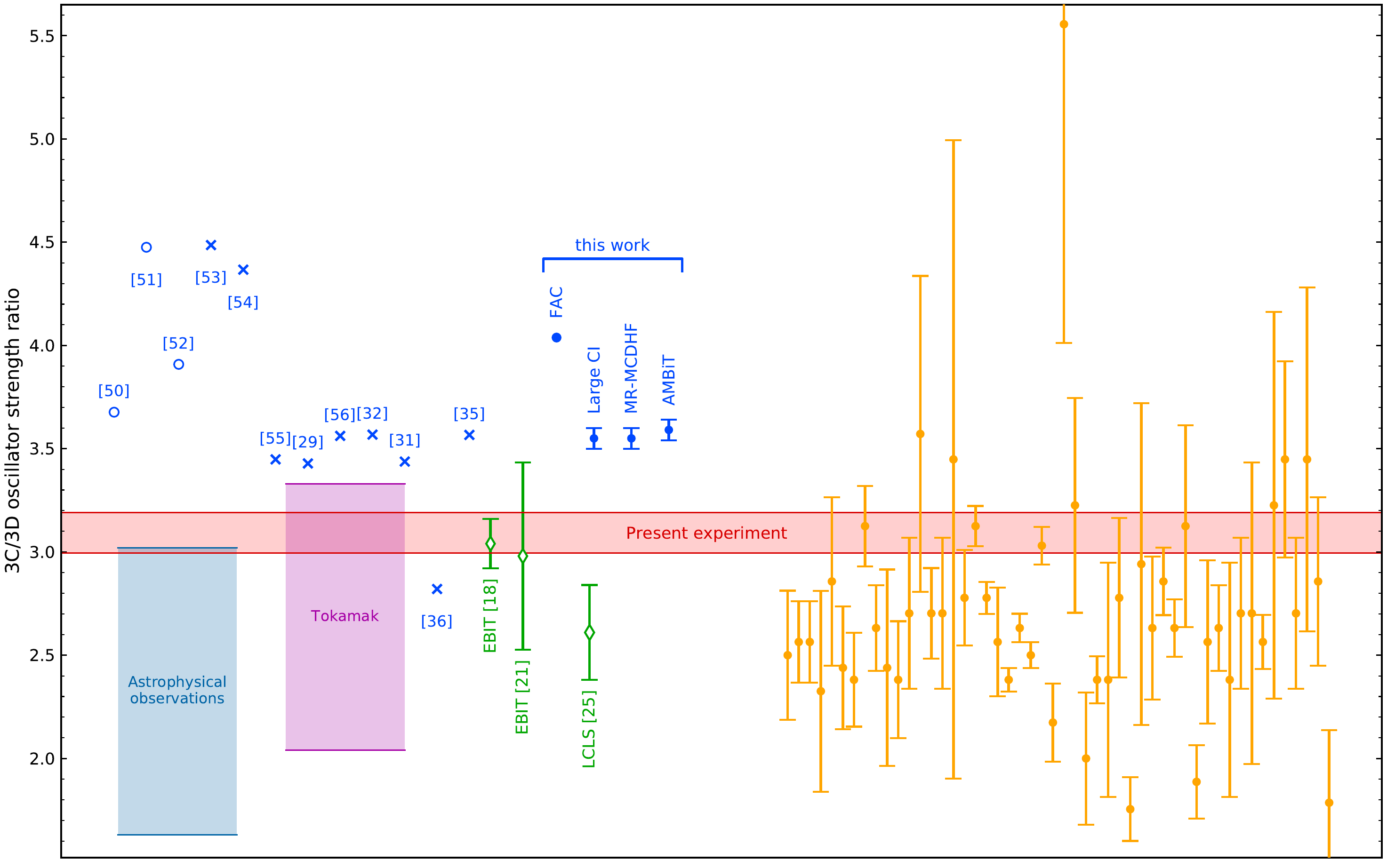}
	\caption{{Present experimental \LINE{3C}/\LINE{3D} ratios -- combined results of three different methods (red band), compared with previous predictions and experiments. Blue open circles: Values from spectral databases. Blue crosses: theoretical predictions. Blue solid circles: our present FAC~\cite{S_shah2019,S_gu2011fac}, large-scale CI (Sec.~\ref{sec:Marianna}), MR-MCDHF (Sec.~\ref{MCDHF})), and AMBiT (Sec.~\ref{ambit}) calculations. Open green diamonds: previous EBIT results. Purple band: spread of Tokamak results~\cite{S_bbg2004}.
			Light blue band: range of ratios observed in the Sun~\cite{S_blake1965,S_mckenzie1980}, Capella~\cite{S_mrd2001,S_behar2001}, and NGC4636~\cite{S_xpb2002}.  Orange solid circles: 46 astrophysical observations of 26 stellar coronae~\cite{S_ness2003} sampled by the RGS onboard \textit{XMM-Newton} and the LETGS/HETGS onboard \textit{Chandra} observatories. Note that the numbers in the square bracket corresponds to the references cited in the main paper.}}
	\label{fig:result_extended}
\end{figure}

\subsection{Estimation of photon peak intensity on sample at P04 PETRA III}

For an estimation of the photon beam intensity on the plasma sample, we refer to the technical parameters of beamline P04 \cite{S_viefhaus2013} and generously assume a total time-averaged photon flux after monochromatization of the beam of $\Phi_{\mathrm{Beam}} = \SI{4e12}{photons\per\second},$ a focal spot size of $\SI{1e-6}{\square\centi\meter}$, and a photon energy of \SI{825}{\electronvolt}. 
The \SI{16}{\nano\second} pulse separation ensures that the upper state population has sufficient time to completely relax into the ground state in between the pulses. We obtain a total energy per bunch of
\begin{equation}
	E_{\mathrm{bunch}} \approx \SI{8.5e-6}{\joule\per\square\centi\meter}.
\end{equation}
Combined with the typical pulse duration of $\SI{44}{\pico\second}$ given by
the official PETRA III datasheet~(see~\footnote{\url{https://photon-science.desy.de/facilities/petra_iii/beamlines/p04_xuv_beamline/unified_data_sheet_p04/index_eng.html}}~and~\footnote{\url{https://photon-science.desy.de/facilities/petra_iii/beamlines/p04_xuv_beamline/beamline_parameters/index_eng.html}}), we deduce an average intensity per pulse of
\begin{equation}
	I \approx \SI{1.9e5}{\watt\per\square\centi\meter}.
\end{equation}
For typical pulse shapes, the peak intensity can be expected to be well below \SI{1e6}{\watt\per\square\centi\meter}.

\newpage

\section{Calculations of 3C and 3D Oscillator Strengths}

\subsection{\label{sec:Marianna} Very large-scale CI calculations}

We start from the solution of the  Dirac-Hartree-Fock equations in the central field approximation to construct the one-particle orbitals. The calculations are carried out using a configuration interaction (CI) method, correlating all 10 electrons. The Breit interaction is included in all calculations.  The QED effects are included following Ref.~\cite{S_QED}. The basis sets of increasing sizes are used to check for convergence of the values. The basis set is designated by the highest principal quantum number for each partial wave included. For example, $[5spdf6g]$ means that all orbitals up to $n=5$ are included for the $spdf$ partial waves and $n=5,6$ orbitals are included for the $g$ partial waves. We find that the inclusion of the $6,7h$ orbitals does not modify the results of the calculations and omit higher partial waves.
The CI many-electron wave function is obtained as a linear
combination of all distinct states of a given angular momentum $J$ and parity
\cite{S_DzuFlaKoz96}:
\begin{equation}
	\label{eq:CIexp}
	\Psi_{J} = \sum_{i} c_{i} \Phi_i\,.
\end{equation}
The energies and wave functions are
determined from the time-independent multiparticle Schr\"odinger equation
$
H \Phi_n = E_n \Phi_n.
$

\begin{table*}[b]
	\centering
	\caption{\label{table1} Contributions to the energies of Fe$^{16+}$ calculated with increased size basis sets and a number of configurations. The results are compared with experiment. All energies are given in cm$^{-1}$ with exception of the last line that shows the difference of the 3C and 3D energies in eV. The basis set is designated by the highest quantum number for each partial wave included. For example, $12spdfg$ means that all orbitals up to $n=12$ are included for $spdfg$ partial waves. Contributions from triple excitations, excitations from the $1s^2$ shells, and QED contributions are given separately.}
	\begin{ruledtabular}
		\begin{tabular}{lccccccccccccc}
			\multicolumn{2}{c}{Configuration}&
			\multicolumn{1}{c}{Expt.~\cite{S_NIST}}&  \multicolumn{1}{c}{Expt.~\cite{S_AK}}&
			\multicolumn{1}{c}{$[5spdf6g]$}&\multicolumn{1}{c}{Triples}& \multicolumn{1}{c}{$1s^2$}&
			\multicolumn{1}{c}{$+[12spdfg]$}& \multicolumn{1}{c}{$+[17dfg]$}&     \multicolumn{1}{c}{QED}&
			\multicolumn{1}{c}{Final}& \multicolumn{1}{c}{Diff.~\cite{S_NIST}}&\multicolumn{1}{c}{Diff.~\cite{S_AK}}& \multicolumn{1}{c}{Diff.~\cite{S_AK}} \\
			\hline
			$2p^6   $&$^1S_0$    &       0&  0       &        0 &   0   &   0&    0 & 0   & 0   & 0       & 0   & 0    &        \\
			$2p^5 3p$&$^3S_1$    & 6093450&  6093295 &  6087185 &   6   & 254& 3876 & 772 & 67  & 6092159 & 1291& 1136 & 0.02\% \\
			$2p^5 3p$&$^3D_2$    & 6121690&  6121484 &  6116210 &  -21  &  24& 2886 & 701 & 43  & 6119842 & 1848& 1642 & 0.03\% \\
			$2p^5 3p$&$^3D_3$    & 6134730&  6134539 &  6129041 &  -23  &  25& 3015 & 711 & 94  & 6132864 & 1866& 1675 & 0.03\% \\
			$2p^5 3p$&$^1P_1$    & 6143850&  6143639 &  6138383 &  -11  &  41& 2825 & 704 & 82  & 6142025 & 1825& 1614 & 0.03\% \\ [0.5pc]
			$2p^5 3s $ &$  2  $  & 5849490&  5849216 &  5842248 & -10   & 108& 3408 & 735 & 787 & 5847276 & 2214& 1940 & 0.03\%  \\
			$2p^5 3s $ &$  1  $  & 5864770&  5864502 &  5857770 & -10   & 70 & 3303 & 708 & 784 & 5862626 & 2144& 1876 & 0.03\%  \\
			$2p^5 3s $ &$  1  $  & 5960870&  5960742 &  5953697 & -10   & 74 & 3364 & 717 &1042 & 5958883 & 1987& 1859 & 0.03\%  \\
			$2p^5 3d $ &$^3P_1$  & 6471800&  6471640 &  6466575 & -11   & 16 & 2384 & 665 & 87  & 6469717 & 2083& 1923 & 0.03\%  \\
			$2p^5 3d $ &$^3P_2$  & 6486400&  6486183 &  6481385 & -13   & 16 & 2250 & 658 & 86  & 6484383 & 2017& 1800 & 0.03\%  \\
			$2p^5 3d $ &$^3F_4$  & 6486830&  6486720 &  6482549 & -12   & 27 & 1745 & 622 & 97  & 6485028 & 1802& 1692 & 0.03\%  \\
			$2p^5 3d $ &$^3F_3$  & 6493030&  6492651 &  6488573 & -14   & 26 & 1740 & 607 & 84  & 6491016 & 2014& 1635 & 0.03\%  \\
			$2p^5 3d $ &$^1D_2$  & 6506700&  6506537 &  6502481 & -17   & 21 & 1696 & 627 & 88  & 6504895 & 1805& 1642 & 0.03\%  \\
			$2p^5 3d $ &$^3D_3$  & 6515350&  6515203 &  6511163 & -18   & 18 & 1762 & 604 & 87  & 6513617 & 1733& 1586 & 0.02\%  \\
			$2p^5 3d $ &$^3D_1$  & 6552200&  6552503 &  6548550 & -16   & -3 & 1747 & 616 & 134 & 6551029 & 1171& 1474 & 0.02\%  \\
			$2p^5 3d $ &$^3F_2$  & 6594360&  6594309 &  6589977 & -16   & 22 & 1729 & 629 & 335 & 6592676 & 1684& 1633 & 0.02\%  \\
			$2p^5 3d $ &$^3D_2$  & 6600950&  6600998 &  6596316 & -17   & 14 & 1947 & 641 & 334 & 6599235 & 1715& 1763 & 0.03\%  \\
			$2p^5 3d $ &$^1F_3$  & 6605150&  6605185 &  6600744 & -17   & 19 & 1803 & 610 & 343 & 6603501 & 1649& 1684 & 0.03\%  \\
			$2p^5 3d $ &$^1P_1$  & 6660000&  6660770 &  6656872 & -8    & -52& 1743 & 619 & 288 & 6659462 & 538 & 1308 & 0.02\%  \\
			3C-3D&      &  13.3655 & 13.4234  & 13.4302&0.0009&-0.0061&-0.0005&0.0004&    0.0191&    13.4440&-0.0785&    -0.0206& 0.15\% \\
		\end{tabular}
	\end{ruledtabular}
\end{table*}

We start with all possible single and double excitations to any orbital up to $5spdf6g$ from the $2s^2 2p^6$, $2s^2 2p^5 3p$  even and
$2s^2 2p^5 3s$, $2s^2 2p^5 3d$, $2s 2p^6 3p$, $2s^2 2p^5 4d$, $2s^2 2p^5 5d$ odd configurations, correlating 8 electrons. We verified that inclusion of the
$2s 2p^6 3s$, $2s^2 2p^5 4f$, $2s^2 2p^5 5f$ even and $2s 2p^6 4p$, $2s^2 2p^5 4s$, and $2s^2 2p^5 5s$ odd configurations as basic configurations have negligible effect on either energies of relevant matrix elements.

The only unusually significant change in the ratio, by 0.07, is due to the inclusion of the $2s^2 2p^3 3d^3$ and $2p^5 3d^3$ configurations.
These are obtained as double excitations from the  $2s^2 2p^5 3d$ odd configuration, prompting the inclusion of the $2s^2 2p^5 4d$, $2s^2 2p^5 5d$ to the list of the basic configurations.

Contributions to the energies of Fe$^{16+}$ calculated with different size basis sets and a number of configurations are listed in Table~\ref{table1}. The results are compared with experimental data from the NIST database \cite{S_NIST} and from a revised analysis
of the experimental data \cite{S_AK}. We use LS coupling and NIST data term designations for comparison purposes, but note that $jj$ coupling would be more appropriate for this ion.
Contributions to the E1 reduced matrix elements $D(3D)=D(2p^6$~$^1S_0 - 2p^5 3d$~$^3D_1)$ and $D(3C)=D(2p^6$~$^1S_0 -2p^5 3d$~$^1P_1)$ and the ratio of the respective oscillator strengths
$$R=\left(
\frac{D(3C)}{D(3D)}\right)^2\times \frac{\Delta E(3C)}{\Delta E(3D)}$$ are listed in  Table~\ref{table2}. The energy ratio is
1.01655.

\begin{table}[t]
	\caption{\label{table2} Contributions to the $E1$ reduced matrix elements $D(3D)=D(2p^6$~$^1S_0 - 2p^5 3d$~$^3D_1)$ and $D(3C)=D(2p^6$~$^1S_0 -2p^5 3d$~$^1P_1)$ (in a.u.) and the ratio of the respective oscillator strengths $R$. See caption of Table~\ref{table1} for designations. $L$ and $V$ rows compared results obtained in length and velocity gauges for the $[12spdfg]$ basis. All other results are calculated using the length gauge. Transition rates are listed in the last row in s$^{-1}$. }
	\begin{ruledtabular}
		\begin{tabular}{lcccc}
			\multicolumn{2}{c}{}&
			\multicolumn{1}{c}{$D(3C)$}&
			\multicolumn{1}{c}{$D(3D)$}& \multicolumn{1}{c}{Ratio}\\
			\hline
			$[5spdf6g]$   &            &0.33492  &  0.17842  &  3.582  \\
			$[5spdf6g]$   &  $+$Triples&0.33493  &  0.17841  &  3.583  \\
			&   Triples  &0.00001  & -0.00001  &    \\
			$[5spdf6g]$   &  +$1s^2$   &0.33480  &  0.17849  &  3.577  \\
			&   $1s^2$   &-0.00012 &  0.00007  &    \\
			$[12spdfg]$ &     $L$      &0.33527  &  0.17884  &  3.573  \\
			&     $V$    &0.33551  &  0.17894  &  3.574   \\
			$+[12spdfg]$&              &0.00036  &  0.00042  &    \\
			$+[17dfg]$    &            &-0.00001 &  0.00001  &    \\
			QED           &            &-0.00017 &  0.00030  &     \\
			Final         &            & 0.33498 &    0.17921  &3.552   \\
			Recomm.       &            &         &           &  3.55(5)   \\
			Transition rate &          &  2.238$\times10^{13}$ &6.098$\times10^{12}$&                \\
		\end{tabular}
	\end{ruledtabular}
\end{table}

\begin{table}[b]
	\centering
	\caption{\label{tab1} Contributions to the 3C and 3D line strengths $S$ and the 3C/3D oscillator strength ratios (energy ratio 1.01655 is used). Energies in eV,  transition rates $A$ in  s$^{-1}$ and natural linewidths $\Gamma$ in meV are listed in the last three rows of the tables.  }
	\begin{ruledtabular}
		\begin{tabular}{lccc}
			
			\multicolumn{1}{c}{}&
			\multicolumn{1}{c}{$S(3C)$}&
			\multicolumn{1}{c}{$S(3D)$}& \multicolumn{1}{c}{Ratio}\\
			\hline
			Small basis              &0.11217  &   0.03183   &  3.582\\
			Medium basis             &0.11241  &   0.03198   &  3.573\\
			Large basis              &0.11240  &   0.03199   &  3.572\\
			+ triple excitations     &0.11241  &   0.03198   &  3.573\\
			+$1s^2$ shell excitations&0.11233  &   0.03201   &  3.567\\
			+QED                     &0.11221  &   0.03212   &  3.552\\
			Final                    &0.1122(2)&   0.0321(4) &  3.55(5)\\
			\hline
			Energies   (eV)              &  825.67 &    812.22   &\\
			$A$ (s$^{-1}$)       & $2.238(4)\times10^{13}$  &   $6.10(7)\times10^{12}$&\\
			$\Gamma$  (meV)      &14.74(3)  &    4.02(5)  &\\
			
		\end{tabular}
	\end{ruledtabular}
\end{table}

We include a very wide range of configurations obtained by triple excitations from the basic configurations as well as excitations from the $1s^2$ shell and find negligible corrections
to both energies and matrix elements as illustrated by Tables~\ref{table1} and \ref{table2}. These contributions are listed as ``Triples'' and ``$1s^2$'' in both tables. A significant increase of the basis set from $[5spdf6g]$ to $[12spdfg]$ improves the agreement of energies with experiment but gives a very small, -0.009, contribution to the ratio. We find that the weights of the configurations containing $12fg$ orbitals are several times higher than those containing $12spd$ orbitals, so we expand the basis
to include more $dfg$ orbitals. We also include $2s^2 2p^3  nd^3$ and $2p^5  nd^3$ configurations up to $n=14$.
The contributions to the energies of the orbitals with $n=13-17$ are $3-5$ times smaller than those with $n=6-12$, clearly showing the convergence of the values with the increase of the basis set. The effect on the ratio is negligible. The uncertainty of the NIST database energies, 3000~cm$^{-1}$ is larger than our differences with the experiment. The energies from the revised analysis of Fe$^{16+}$ spectra ~\cite{S_AK} are estimated to be accurate to about~90~cm$^{-1}$
and the scatter of the differences of different levels with the experiment is reduced. The last line of Table~\ref{table1} shows the difference of the $3C$ and $3D$ energies in eV, with the final value 13.44(5)eV.
We explored
several different ways to construct the basis set orbitals. While the final results with an infinitely large basis set and complete configurations set should be identical, the convergence properties of the different basis sets vary, giving about 0.04 difference in the ratio and 0.04~eV in the $3C-3D$ energy difference at the $12spdfg$ level.
Therefore, we set an uncertainty of the final value of the ratio to be $0.05$.  As an independent test of the quality and completeness  of the current basis set, we compare the results for  $D(3C)$ and $D(3D)$ obtained in length and velocity gauges for the $[12spdfg]$ basis, see  rows
$L$ and $V$ in Table~\ref{table2}. The difference in the results is only 0.001. 
The final results for the line strengths $S$ and the 3C/3D oscillator strength ratio after several stages of computations are summarised in Table~\ref{tab1}, which clearly illustrates a very small effect of all corrections. 

This work was supported in part by U.S. NSF Grant No.\ PHY-1620687 and RFBR grants No.\ 17-02-00216 and No.\ 18-03-01220.

\subsection{\label{MCDHF} Multiconfiguration Dirac-Hartree-Fock calculations}

In the multiconfiguration Dirac-Hartree-Fock (MCDHF) method, similarly to the CI approach outlined in the previous section, the many-electron state is given
as an expansion in terms of a large set of $jj$-coupled configuration state functions [see Eq.~(\ref{eq:CIexp})]. In contrast to the CI
calculations, in the case of MCDHF, the single-electron wave functions (orbitals) are self-consistently optimized. We use the method in one of its most recent implementations,
namely, applying the \textsc{GRASP2018} code package~\cite{S_GRASP2018}. For the virtual orbitals, the optimization of the orbitals was done in a layer-by-layer approach,
i.e. when adding a new layer of orbitals (in our case, orbitals in the same shell) in the configuration expansions, the lower-lying single-electron functions are kept frozen.

\begin{figure}[b]
	\includegraphics[width=1.08\textwidth]{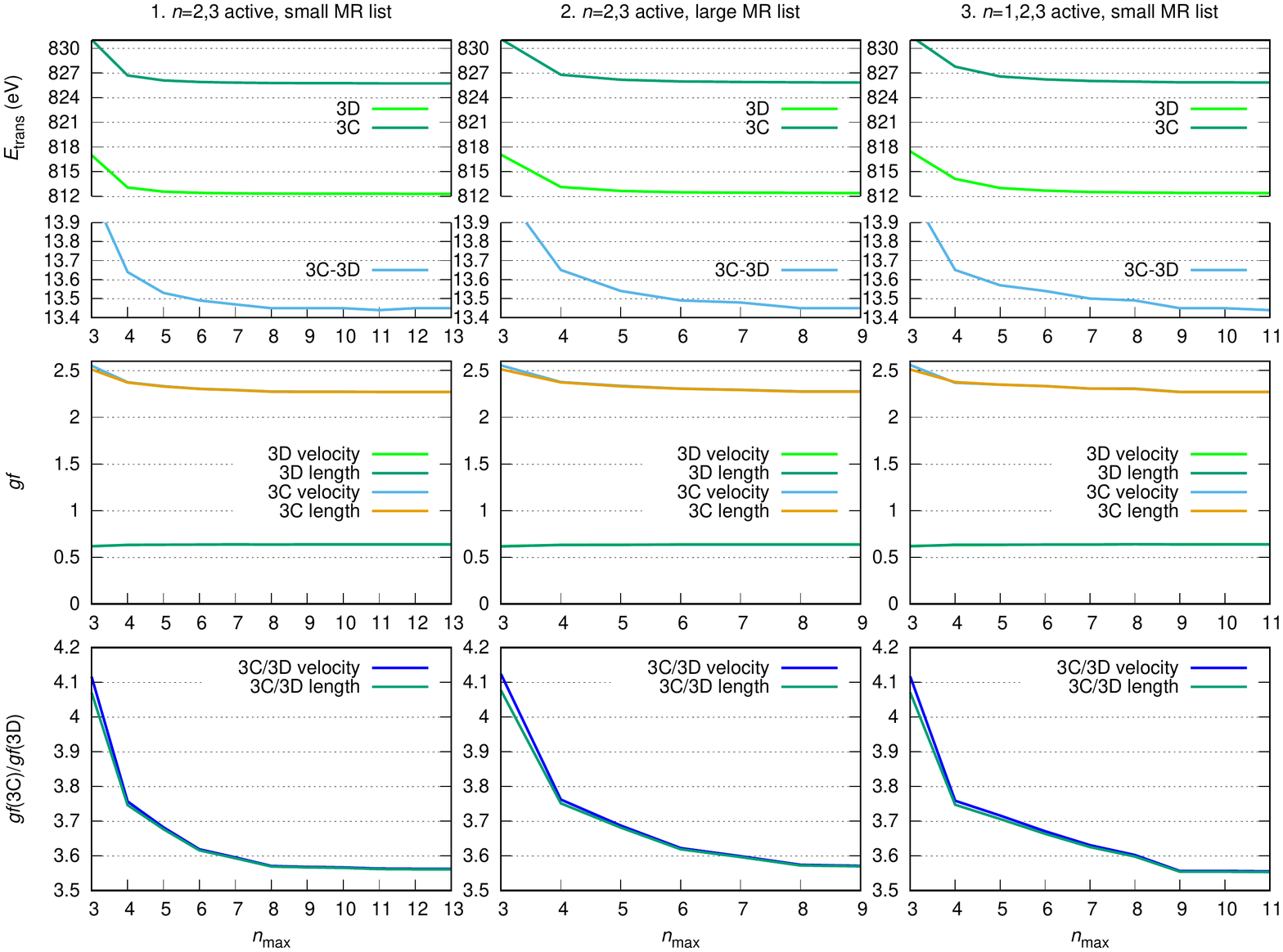}
	\caption{
		\label{fig:1}
		Convergence of the MCDHF calculations: The X-ray transition energies of the 3C and 3D lines, their difference, the weighted oscillator strengths $gf$ 
		and their ratio vs. the maximal principal quantum number $n_{\rm max}$ used.
		Oscillator strengths are given both in the relativistic length (Babushkin) and velocity (Coulomb) gauges to numerically control gauge invariance.
		The different columns display results from different sets of calculations, as described in the text.
	}
\end{figure}

In a first set of calculations, we use the $2s^2 2p^6$ configuration for the ground state and the $2s^2 2p^5 3s$, $2s^2 2p^5 3d$, and $2s^1 2p^6 3p$ $J=1$ odd configurations for the excited states
to generate the configuration lists. Single and double electron exchanges from the $n=2,3$ spectroscopic (occupied) orbitals were taken into account up to virtual orbitals $n_{\rm max}spdfg$,
where the maximal principal quantum number $n_{\rm max}$ is varied in the computations to study the convergence of the results. Such an approach is helpful in estimating the final
theoretical uncertainty. Test calculations also using virtual orbitals with $h$ and $i$ symmetry have shown that these high angular momenta do not play a noticeable role.
The ground and excited states were treated separately, i.e. two independently optimized sets of orbitals were used.
After these multireference MCDHF calculations, the possible effects of further higher-order electron exchanges were included in a subsequent step, when a CI calculation
was performed with the extended configuration lists (triple excitations from the multireference states up to $n=4$ orbitals, yielding approx. 800 thousand configurations for $n_{\rm max}=13$),
employing the radial wave functions obtained from the previous MCDHF calculations. Furthermore, the effects of the frequency-independent Breit relativistic electron interaction operator, the normal
and specific mass shift, and approximate radiative corrections are accounted for (see \cite{S_GRASP2018} and references therein). The QED effects have been included in the calculation of transition
energies (which also enter the oscillator strengths), however, not in the electric dipole matrix elements, as such corrections are anticipated to be on the 1\% level and thus can be neglected. The oscillator strengths were evaluated
with the biorthogonal basis sets, each optimized separately for the ground- and excited states, to include orbital relaxation effects. The results of these calculations are presented in the first
column of Fig.~\ref{fig:1}. The bottom panel shows that the oscillator strength ratio is converged from $n_{\rm max}=9$ on.

In a subsequent set of calculations, the multireference set describing the ground and excited levels were expanded to include all $J=0$ even and $J=1$ odd states with 1 or 2 electrons in the M shell.
The maximal principal quantum number of the virtual orbitals was set to 9 to limit the computational expense of calculations. Results are shown in the 2nd column of Fig.~\ref{fig:1}.
In a third setting, calculations were performed with the smaller
multireference list as described in the previous paragraph, however, with all spectroscopic orbitals (those with $n=1,2,3$) included in the active set of orbitals when generating the configuration
list. With the triple electron exchanges also included, this procedure yielded approx. 1.2 million configurations in the description of the excited states. The energies and strengths are shown in the last column
of the figure. The converged 3C/3D oscillator strength ratios agree well for all 3 calculations. Comparing the different results, the final value for the ratio is 3.55(5), which agrees well
with earlier large-scale MCDHF results~\cite{S_sbr2012,S_jonsson2014}, and also with the results of the other theoretical methods described in this Supplement. For the difference of the energies of the 3C
and 3D lines -- which can be more accurately determined in the experiment than the absolute X-ray transition energies --  we obtain 13.44(5)~eV.

\subsection{\label{ambit}\ambit: particle-hole CI method calculations}

A separate CI calculation of the 3C and 3D lines in Fe$^{16+}$ has been performed with the \ambit\ code~\cite{S_kahl19cpc}. Our calculation begins with a Dirac-Hartree-Fock calculation of Ne-like Fe to construct the core $1s$, $2s$, and $2p$ orbitals in the $V^{N}$ potential. The Breit interaction is included throughout the calculation. We diagonalize a set of $B$-splines in the Dirac-Fock potential to obtain valence orbitals. Configuration interaction is performed using the particle-hole CI method~\cite{S_berengut16pra}, however, this can be mapped exactly to the electron-only approach described in Section~\ref{sec:Marianna}.

Our basic calculation is presented on the first line of Table~\ref{tab:ambit}. The CI space consists of all possible single and double excitations up to $10spdf$ from the same set of leading configurations presented previously: $2s^2\ 2p^6$, $2p^{-1}\ 3p$, $2p^{-1}\ 3s$, $2p^{-1}\ 3d$, $2p^{-1}\ 4d$, $2p^{-1}\ 5d$, and $2s^{-1} 3p$.
At this stage, we do not include excitations from the frozen $1s^2$ core. Even for this calculation, the matrix size for the odd-parity $J = 1$ levels is $N = 479\,075$. To reduce the number of stored matrix elements we use emu CI~\cite{S_geddes18pra0}, where interactions between high-lying configuration state functions are ignored. We limit the smaller side of the matrix to only including double excitations up to $5spdf$ and limit the number of $2s$ and $2p$ holes to single removals from an expanded set of leading configurations which include, in addition to those listed above, $2p^{-2}\ 3d^2$, $2p^{-1}\ 2s^{-1}\ 3d^2$, and $2s^{-2}\ 3d^2$. This results in a reduced small side $N_{small} = 80497$. We have checked that expanding the configuration state functions included in $N_{small}$ makes no difference to our results at the displayed accuracy.

All of our calculations include the Breit interaction at all stages, and the dipole matrix elements are calculated in the relativistic formulation with $\omega = 30$~a.u. In the second and third lines of Table~\ref{tab:ambit}, we show the effects of removing the Breit interaction and using the static dipole matrix element ($\omega = 0$), respectively.

We then expand our calculation to include $g$-wave excitations, up to basis $10spdfg$. The difference from $10spdf$ is shown in the fourth line of Table~\ref{tab:ambit}. We see in the \ambit\ calculation very little effect from the inclusion of these waves. In the fifth line, we show the effect of allowing excitations from $1s^2$, and in the sixth line we see the effect of including the Uehling potential~\cite{S_ginges16jpb} and self-energy~\cite{S_ginges16pra} using the radiative-potential method~\cite{S_flambaum05pra}. This broadly agrees with the model-operator QED presented in Table~\ref{table2}.

The final row of Table~\ref{tab:ambit} gives the results including excitations to $g$-waves, excitations from $1s^2$, and QED effects. The total number of configuration state functions accounted for is over 1.25 million for the odd-parity $J = 1$ symmetry. Nevertheless, the CI is not quite converged with respect to including orbitals with $n > 10$. We estimate based on calculations for $8spdf$ and $12spdf$ that the uncertainty in level energies is conservatively of order $3000~\textrm{cm}^{-1}$ and for the ratio $gf_{3C}/gf_{3D}$ is of order 0.05. These results are consistent with the CI calculation presented in Sec.~\ref{sec:Marianna}
and the MCDHF results presented in Sec.~\ref{MCDHF}.

\begin{table*}[!htb]
	\caption{\label{tab:ambit} Particle-hole CI calculations using \texttt{AMBiT} of the level energies $E_{3C}$ and $E_{3D}$ (cm$^{-1}$), reduced matrix elements $D(3C)$ and $D(3D)$ (a.u.), and ratio of oscillator strengths of the $3C$ and $3D$ transitions in Fe$^{16+}$.
	}
	\begin{ruledtabular}
		\begin{tabular}{lccccc}
			& $E_{3C}$ & $E_{3D}$ & $D(3C)$ & $D(3D)$ & $gf_{3C}/gf_{3D}$ \\
			\hline
			[$10spdf$] & 6661400 & 6552424 & 0.33893 & 0.17992 & 3.607 \\
			without Breit & 6668847 & 6557636 & 0.33758 & 0.18164 & 3.511 \\
			$\omega = 0$ &   &  
			& 0.33937 & 0.18009 & 3.610 \\
			+[$10g$] & -15 & -13
			& -0.000001 & 0.000003
			& \\
			+$1s^2$ & -103 & -85 & -0.000040 & 0.000016
			& \\
			+QED & 235 & 94
			& -0.00012 & 0.00031
			& \\
			Final & 6661517 & 6552420
			& 0.33877 & 0.18025 & 3.591 \\
		\end{tabular}
	\end{ruledtabular}
\end{table*}

\end{document}